\documentclass[10pt,a4paper,twosides]{article}

\usepackage{amsmath,amssymb,amsfonts}

\newcommand{\EXP}[1]{\mathrm{e}^{#1}} 
\newcommand{\DEF}{\overset{\mathrm{def}}{=}}
\newcommand{\DEFt}{\overset{\text{\tiny def}}{=}}
\newcommand{\imat}{{\mathrm{i}}} 
\newcommand{\dmat}{\mathrm{d}}
\newcommand{\diag}{\mathrm{diag}}

\begin{document}
\title{Algebraic spectral gaps}
\author{Amaury Mouchet\\ Laboratoire de Math\'ematiques et Physique Th\'eorique\\ Universit\'e Fran\c{c}ois Rabelais de Tours\\ 
\textsc{\textsc{cnrs (umr 6083)}},
F\'ed\'eration Denis Poisson\\
 Parc de Grandmont 37200
  Tours,  France.\\
 \texttt{mouchet@phys.univ-tours.fr}}\date{September 30, 2011}

\maketitle
\begin{abstract} For the one-dimensional Schr\"odinger equation, some real intervals with
 no eigenvalues  (the spectral gaps) may be obtained rather systematically with a
method proposed by H.~Giacomini and A.~Mouchet in 2007. The present
 article provides some alternative 
formulation of this method, suggests some possible generalizations and 
extensively discusses the higher-dimensional case.
\\
\\   H.~Giacomini et A.~Mouchet ont propos\'e en 2007 une m\'ethode permettant d'obtenir
des trous spectraux d'un op\'erateur de Schr\"odinger \`a une dimension, c'est-\`a-dire des intervalles ne contenant aucune valeur propre. Le pr\'esent article introduit une formulation
diff\'erente de cette m\'ethode, sugg\`ere des g\'en\'eralisations possibles et traite
de fa\c{c}on exhaustive le cas de plusieurs dimensions. 
\end{abstract}

\section*{Introduction}

A spectral gap (or eigengap) of a self-adjoint operator is a closed real
interval to which no eigenvalues belongs. 
In \cite{Giacomini/Mouchet07a} we have presented a systematic method for 
finding gaps in the discrete part of the spectrum of a one-dimensional  
non-magnetic Schr\"odinger equation with a potential $V (x)$. 
When $V$ is a polynomial half-bounded from below, the boundaries
of the gaps are given by the real zeroes of a family of polynomials
whose degree~$D$ may be arbitrary large. The
construction of these polynomials is provided by an explicit and straightforward
algorithm. For still not understood reasons, it happens that in every 
case we have considered, our method works surprisingly well when compared to numerical computations: when increasing~$D$ 
the more and more numerous intervals
we compute resolve the spectrum from below (i.e. the lowest eigenvalues are separated
by at least one gap) and the infima of each interval seem 
to converge quickly to the eigenvalues.
For the moment, we have no clue to understand these two
 phenomena and our method comes 
without any estimation of the distance between the gaps and the spectrum.

Being as local as possible (no computation of integrals is required),
our method differs strongly in spirit from other 
spectral approximations like the Rayleigh-Ritz variational
methods or the Rayleigh-Schr\"odinger perturbation methods. 
Since generically the spectrum cannot
be determined exactly, finding such gaps may offer a valuable piece
of spectral information, complementary to the information 
obtained by other methods.

One natural issue is to try to extend our method to
multi-dimensional systems.  Then, the nature of the spectrum
depends on the integrability properties of the
system (among the vast literature on this subject see for 
instance \cite{Haake01a}). Qualitatively  the statistical distribution of the 
eigenvalues exhibits different correlations according to the nature of its symmetries.
For instance, for non-integrable systems, the two point correlation function 
of the discrete spectrum
exhibits a so-called ``level repulsion'' because, unlike what
 occurs in integrable cases, the probability of finding 
 two successive eigenvalues
whose distance is $s$ vanishes  when $s$ tends to zero. Therefore we expect
that this dichotomy between integrable and non-integrable cases should somehow appear
in any general method for finding spectral gaps. However, most unfortunately,
we have not been able to generalise our strategy in higher dimensions.
Although there was \emph{a priori} no obstacle in sight to such an attempt, it happened
that the origin of the obstruction came from very subtle arguments
that are deeply hidden. One aim of this note is to explain (in section~\ref{sec:anyd}) this negative result 
with the hope that it may help to find out some way to bypass the pitfalls or, at least,
to help avoiding the same tracks. 

Our second aim is more optimistic but still rely on speculative 
grounds. 
After recalling the main ingredients of our method in section~\ref{subsec:principle},
I will introduce a systematic algebraic approach  which at first sight
seems to rephrase in a more elaborate way what we have done in
\cite{Giacomini/Mouchet07a}. However by associating with the Schr\"odinger equation
the closed algebra of differential operators that will be introduced in~\S\S~\ref{subsec:alg_i}, 
\ref{subsec:closedalg}, we can easily guess
a fruitful strategy to deal with spectral problems associated with more general (1d) linear equations --- for instance
of order larger than two --- or even with non-linear equation like the Gross-Pitaevskii equation ($g$ is a real coupling constant)
\begin{equation}\label{eq:grosspitaevskii}
  \varphi''=(V-E)\varphi+g\varphi^3\;.
\end{equation}

\section{The one dimensional case: an algebraic approach}\label{sec:1d}

\subsection{The principle of the method}\label{subsec:principle}

The stationary one-dimensional non-magnetic Schr\"odinger equation can be written as follows
\begin{equation}\label{eq:schrodinger1d}
  \frac{\dmat^2\varphi}{\dmat x^2}=2(V(x)-E)\varphi\;.
\end{equation}
where we will take the real potential~$V$ to be smoothly defined on~$\mathbf{R}$. 
The real~$E$ will be an eigenvalue 
whenever the real function $\varphi$  is square integrable on~$\mathbf{R}$.  
The key idea of our method is to construct, for a given integer~$N\geqslant1$, a
real function~$J_N(\varphi',\varphi,x,E)$
(the prime stands for the derivative~$\dmat/\dmat x$) such that
\begin{eqnarray*}
\mathrm{(i)} &&\frac{\dmat}{\dmat x}\Big(J_N\big(\varphi'(x),\varphi(x),x,E\big)\Big)
               =\big(\varphi(x)\big)^N \, F_{N}(x,E)\;,\\
\mathrm{(ii)} &&\lim_{|x|\to+\infty}J\big(\varphi'(x),\varphi(x),x,E\big)=0.
\end{eqnarray*}
The real function~$F_N$ of both the spatial
coordinate~$x$ and the energy~$E$ is 
$\varphi$\textit{-independent} and is obtained from the
potential~$V$ and its derivatives. 
For instance we will find the following expressions:
\begin{subequations}\label{eq:FN}
\begin{align}
 \label{eq:F1}F_1=&  -a''_0+2(V-E)\,a^{}_0\;, \\
 \label{eq:F2}F_2=&\frac{1}{2}\,a'''_0-4(V-E)\,a'_0-2V'a^{}_0\;,\\
 \label{eq:F3}F_3=&-\frac{1}{6}\,a^{(iv)}_0+\frac{10}{3}\,(V-E)\,a''_0+\frac{10}
{3}\,V'a'_0
                   +\left(\,V''-6\,(V-E)^2\right)a^{}_0\;,\\
              F_4=&\frac{1}{24}\,a^{(v)}_0-\frac{5}{3}\,(V-E)\,a'''_0 -\frac{5}{
2}\,V'a_0''
                   +\left(-\frac{3}{2}\,V''+\frac{32}{3}\,(V-E)^2\right)a'_0\nonumber\\
 \label{eq:F4}    &+\left(-\frac{1}{3}\,V'''+\frac{32}{3}\,V'(V-E)\right)a^{}_0\;.
\end{align}
\end{subequations}
where~$a_0$ is any smooth function such that
\begin{equation}
  \lim_{|x|\to\infty}a_0(x)|\varphi|^N=0
\end{equation}
 which is not very restrictive since we know using semi-classical arguments that $\varphi$ itself is exponentially decreasing at infinity 
\cite[chap.10]{Bender/Orszag78a}:
\begin{equation}
  \varphi(x)\underset{|x|\to\infty}{\sim}\EXP{-\int^{|x|}\sqrt{2(V(x)-E)}\,\dmat x}
\end{equation}

 Condition~(i) is the cornerstone of our method and,
before justifying how it can be obtained (we will see that condition~(ii) is not so
restrictive), let us first explain how 
gaps in the spectrum may be obtained.  

When the conditions~(i) and (ii) are simultaneously
fulfilled, an immediate consequence is that the integral~$\int_{-\infty}^{+\infty}
\big(\varphi(x)\big)^N \, F_N(x,E)\, \dmat x$ vanishes. This implies
that, if $E$ is truly an eigenenergy, the function
$x\mapsto\big(\varphi(x)\big)^N \, F_N(x,E)$ should change its sign.
If~$N$ is even, we obtain a $\varphi$-independent condition:
\textit{for any fixed energy $x\mapsto F_N(x,E)$ must change its sign
on the real axis}.  For such a
one-dimensional problem, and for a given~$N$, we still can choose~$F_N$
in a wide continuous set of smooth functions on the real axis because
we have a lot of freedom in choosing~$a_0$. A forbidden value of~$E$
(\textit{i.e.} $E$ cannot be an eigenenergy) is
obtained if we are able to chose~$a_0$ such that~$F_N$ remains positive on
the whole $x$-axis. Once this property is achieved, it remains stable
 under small perturbations within the set of~$a_0$'s, for instance by varying
the control parameters~$\lambda$ on which~$a_0$ may depend,
and we obtain
a whole interval where no eigenenergy can exist. More precisely, if we 
introduce explicitly the $\lambda$-dependence in $F_N$, the boundaries of the gaps will
necessary be given by some solutions of the system of equations
\begin{subequations}\label{subeq:equationsforthegaps}
\begin{align}  
  F_N(x,E,\lambda)&=0\;;\label{eq:Fdequals0} \\
  \partial_x F_N(x,E,\lambda)&=0\;;\label{eq:Fdx}\\
   \partial_\lambda F_N(x,E,\lambda)&=0\;.\label{eq:Fdlambda}
\end{align}
Using the implicit function theorem  where
\begin{equation}
  \left|\begin{matrix}\partial_E F_N & \partial^2_{xE}F_N   \\ 
                    \partial_{x}F_N         &\partial^2_{xx}F_N
\end{matrix}\right|=|\partial_EF_N| \, |\partial^2_{xx}F_N|\neq0\;,\label{eq:nondegenerate}
\end{equation}
\end{subequations} 
the first two equations define implicitly $x(\lambda)$ and~$E(\lambda)$; these are the 
 conditions for a bifurcation in the zeroes of~$x\mapsto F_N(x,E,\lambda)$ to occur.
 On one side of the bifurcation~$x\mapsto F_N(x,E,\lambda)$ has locally a 
constant sign (and therefore the corresponding value~$E(\lambda)$ is forbidden) 
whereas on the other side~$x\mapsto F_N(x,E,\lambda)$ locally changes its sign and~$E(\lambda)$
cannot be ruled out from the spectrum. To put it differently, in the $(x,E,\lambda)$, the
set of zeroes of~$F_N$ becomes tangent to the~$x$-space. Then for each value
of $\lambda$, $E(\lambda)$ is a candidate for being the boundary of a spectral gap. If this is 
the case, we can reach an extremal value 
provided~$0=\partial E/\partial\lambda=-\partial_\lambda F_N/\partial_E F_N$ that leads 
to the equation~\eqref{eq:Fdlambda}.

See \cite{Giacomini/Mouchet07a} for an effective implementation of 
this method and for applications. In the following, we will remain at a more formal level and let us start by defining some notations. 

\subsection{Algebraic construction of condition (i) and classification of the possible~$F_N$'s}\label{subsec:alg_i}

Denote by~$\hat{\mathcal{P}}_N$, the vector space of smooth applications from~$\mathbf{R}$
to~$\mathbf{R}^{N+1}$. Any element $a$ of~$\hat{\mathcal{P}}_N$ may be represented
by a vector 
field~$x\mapsto\big(a_n(x)\big)_{n\in\{0,\dots,N\}}=\big(a_0(x),\dots,a_N(x)\big)$ and can be associated in a one-to-one correspondence with the
 homogeneous polynomial of degree~$N$ in the two 
variables~$\Phi$ and~$\Psi$:
\begin{equation}
  P_a(\Psi,\Phi,x)\DEF\sum_{n=0}^N a_n(x)\Psi^{N-n}\Phi^{n}\;;\end{equation}
then, it may be used to construct the real function on~$\mathbf{R}$ defined by
\begin{equation}
  P_a\big(\varphi'(x),\varphi(x),x\big)
\DEF\sum_{n=0}^N a_n(x)\big(\varphi'(x)\big)^{N-n}\big(\varphi(x)\big)^{n}\;.
\end{equation}
For computations we will distinguish the ``total'' derivative~$\mathsf{D}$
of a function $P\big(\varphi'(x),\varphi(x),x\big)$ from its partial 
derivatives~$\partial_{\varphi'}$, $\partial_\varphi$ and $\partial_x$:
\begin{equation}
  \mathsf{D}P\big(\varphi'(x),\varphi(x),x\big)\DEF 
v(x)\varphi(x)\partial_{\varphi'}P\big(\varphi'(x),\varphi(x),x\big)
+\varphi'(x)\partial_{\varphi}P\big(\varphi'(x),\varphi(x),x\big)+\partial_x P\big(\varphi'(x),\varphi(x),x\big)
\end{equation}
 where the substitution~$\varphi''=v\varphi$ has been made since we
suppose that~$\varphi$ fulfills \eqref{eq:schrodinger1d}. For simplicity
we have left implicit the~$E$-dependence in
\begin{equation}\label{def:v}
  v(x)\DEF2\big(V(x)-E\big).
\end{equation}
From its very definition, it is obvious that the set~$\mathcal{P}_N$ of homogeneous polynomials
 of degree~$N$ in~$\varphi'(x)$ and~$\varphi(x)$ is stable under~$\mathsf{D}$ and, 
moreover, $\mathsf{D}$ is represented by a linear operator $\hat{\mathsf{D}}$ in~$\hat{\mathcal{P}}_N$:
\begin{equation}\label{eq:opD}
\hat{\mathsf{D}}\begin{pmatrix} a_0\\[.4cm]a_1\\[.4cm]a_2\\[.4cm]\vdots\\[.4cm] 
                 a_n\\[.4cm]\vdots\\[.5cm]
                 a_{N-2}\\[.4cm]a_{N-1}\\[.4cm]a_N
\end{pmatrix}
=\begin{pmatrix}\partial_x & 1 & 0 &0 &0 &\cdots&\cdots&\cdots &0\\[.2cm]
                 Nv&\partial_x& 2 &0&0 &\ddots& & &\vdots \\[.2cm]
                 0&(N-1)v&\partial_x& 3&0&\ddots& & &\vdots\\[.2cm]
                 \vdots&\ddots&\ddots&\ddots&\ddots&\ddots&\ \ddots&\ddots&\vdots\\[.3cm]
                 \vdots&\ddots& 0&(N\!-\!n\!+\!1)v&\partial_x\ &n+1\ &0&\ddots&\vdots  \\[.3cm]   
                 \vdots&\ddots&\ddots&\ddots&\ddots&\ddots&\ddots&\ddots&\vdots\\[.3cm]  
                 \vdots&&&\ddots& 0&3v&\partial_x&N-1&0  \\[.2cm]   
                 \vdots&&&\ddots&0&0&2v&\partial_x&N  \\[.3cm]  
                 0&\cdots&\cdots&\cdots&0&0&0&v&\partial_x  \\ 
                         
\end{pmatrix}
 \begin{pmatrix} a_0\\[.4cm]a_1\\[.4cm]a_2\\[.4cm]\vdots\\[.4cm] 
                 a_n\\[.4cm]\vdots\\[.5cm]
                 a_{N-2}\\[.4cm]a_{N-1}\\[.4cm]a_N
\end{pmatrix}\;.
\end{equation}
For each~$n\in\{0,\cdots,N\}$, denote by~$\hat{\mathcal{Q}}_n$ 
the subspace of~$\hat{\mathcal{P}}_N$ defined by~$a_n=0$ and $\hat{\bar{\mathcal{Q}}}_n$ its complementary
defined by the direct sum decomposition $\hat{\mathcal{P}}_N=\hat{\mathcal{Q}}_n\oplus\hat{\bar{\mathcal{Q}}}_n$.  
Looking for all the~$J_N$ that fulfill condition~(i) can therefore be interpreted  as the determination
in~$\hat{\mathcal{P}}_N$ of the preimage~$\hat{\mathsf{D}}^{-1}\hat{\bar{\mathcal{Q}}}_N$.
In \cite{Giacomini/Mouchet07a} we have shown how to
straightforwardly compute~$J_N$  and obtain~$F_N$ but let us propose a strategy based on a more algebraic formalism that may
be useful as a warming up for higher dimensions.

 From any~$a$, we can systematically
reduce the degree in~$\varphi'$ of~$P_a$ if we work up to a total derivative. Indeed, for any
monomial caracterised by~$n\in\{0,\cdots,N-1\}$, we use the identity
\begin{align}
  a^{}_n\varphi'^{N-n}\varphi^{n}&=a^{}_n\varphi'^{N-n-1}D\left(\frac{1}{n+1}\varphi^{n+1}\right)\;;\\ 
&= -\frac{1}{n+1}\,a'_n\,\varphi'^{N-n-1}\varphi^{n+1}-\frac{N-n-1}{n+1}\,va^{}_n\,\varphi'^{N-n-2}\varphi^{n+2}
  +D\left(\frac{1}{n+1}\,a^{}_n\,\varphi'^{N-n-1}\varphi^{n+1}\right),\label{eq:reductionmoduloImD}
\end{align}
where, again, we have substituted~$\varphi''$ by~$v\varphi$. The operation that 
transforms~$a_n\varphi'^{N-n}\varphi^{n}$ to the two first terms in~\eqref{eq:reductionmoduloImD} may be 
linearly represented in~$\hat{\mathcal{P}}_N$ by the reduction operator defined by
\begin{equation}\label{eq:opR}
\hat{\mathsf{R}}\begin{pmatrix} a_0\\[.4cm]a_1\\[.4cm]a_2\\[.4cm]\vdots\\[.4cm] 
                 a_n\\[.4cm]
                 \vdots\\[.4cm]a_{N-1}\\[.4cm]a_N
\end{pmatrix}
\DEF\begin{pmatrix}0 & 0 & 0 &\cdots  &\cdots&\cdots&\cdots &0\\[.2cm]
                 -\partial_x& 0& 0 &\ddots && & &\vdots \\[.2cm]
                 (1-N)v& -\frac{1}{2}\partial_x&\ddots & \ddots&& & &\vdots\\[.2cm]
                 \vdots&\ddots&\ddots&\ddots&\ddots&\ddots&\ddots&\vdots\\[.2cm]
                 \vdots&0& \frac{n-1-N}{n-1}v&-\frac{1}{n}\partial_x &0 &0&\ddots&\vdots  \\[.3cm]   
                 \vdots&\ddots&\ddots&\ddots&\ddots&\ddots&\ddots&\vdots\\[.3cm]   
                 \vdots&&\ddots&0&\frac{-2}{N-2}v&-\frac{1}{N-1}\partial_x&0&0  \\[.3cm]  
                 0&\cdots&\cdots&\cdots&0&\frac{-1}{N-1}v&-\frac{1}{N}\partial_x &1  \\ 
                         
\end{pmatrix}
 \begin{pmatrix} a_0\\[.4cm]a_1\\[.4cm]a_2\\[.4cm]\vdots\\[.4cm] 
                 a_n\\[.4cm]
                 \vdots\\[.4cm]a_{N-1}\\[.4cm]a_N
\end{pmatrix}\;.
\end{equation}
For~$N=1$, we have~$\hat{\mathsf{R}}=\big(\begin{smallmatrix}0 &0 \\ -\partial_x & 1\end{smallmatrix}\big)$.
By construction we have $a-\hat{\mathsf{R}}a=0$ for any 
vector~$a=(0,\dots,0,a_N)\in\hat{\bar{\mathcal{Q}}}_N$; by
\eqref{eq:reductionmoduloImD}, we have  $a-\hat{\mathsf{R}}a\in\mathrm{Im}\hat{D}$ for any 
vector~$a\in\hat{\bar{\mathcal{Q}}}_n$ with~$n\in\{0,\cdots,N-1\}$ ; therefore, by linearity,
we have $a-\hat{\mathsf{R}}a\in\mathrm{Im}\hat{D}$ for any $a\in\hat{\mathcal{P}}_N$\;.
 Moreover, since
$\hat{\mathsf{R}}$ is a lower triangular matrix with all diagonal terms but one being zero, all the components
of $\hat{\mathsf{R}}^N a$ vanish but the last one $A_N[a]\DEF(\hat{\mathsf{R}}^N a)_N$ which is a differential operator
on~$a$ of order~$N$. For instance we have
\begin{subequations}\label{eq:AN}
\begin{align} 
  A_1[a]=&-a_0'+a^{}_1\;;\\
  A_2[a]=&\frac{1}{2}a_0''-va^{}_0-\frac{1}{2}a_1'+a^{}_2\;;\\
  A_3[a]=&-\frac{1}{6}a_0'''+\frac{7}{6}va_0'+\frac{2}{3}v'a^{}_0+\frac{1}{6}a_1''-\frac{1}{2}va^{}_1-\frac{1}{3}a_2'+a^{}_3\;;\\
  A_4[a]=&\frac{1}{24}a_0^{(iv)}-\frac{2}{3}va_0''-\frac{3}{4}v'a_0'-(\frac{1}{4}v''-v^2)a^{}_0-\frac{1}{24}a_1'''
+\frac{5}{12}va_1'+\frac{1}{4}v'a^{}_1+\frac{1}{12}a_2''-\frac{1}{3}va^{}_2-\frac{1}{4}a_3'+a^{}_4\;.
\end{align}
\end{subequations}
Now if we use the decomposition
\begin{equation}
  1=\hat{\mathsf{R}}^N+(1-\hat{\mathsf{R}})\sum_{n=0}^{N-1}\hat{\mathsf{R}}^n\,,
\end{equation}
it can be seen immediately  that any~$a$ can be uniquely decomposed in $\hat{\mathsf{R}}^N a\in\hat{\bar{\mathcal{Q}}}_N$ 
plus a vector in~$\mathrm{Im}\hat{\mathsf{D}}$.
Translating this decomposition into the language of functions and taking for~$a$ the vector associated 
to~$\mathsf{D}K_N\big(\varphi'(x),\varphi(x),x\big)$ where~$K_N$ is any homogeneous polynomial of degree~$N$ in
$\big(\varphi'(x),\varphi(x)\big)$,  we
have shown that there always exists a homogeneous polynomial~$\tilde{K}_N\big(\varphi'(x),\varphi(x),x\big)$ of degree~$N$ in
$\big(\varphi'(x),\varphi(x)\big)$, such that 
\begin{equation}
 \mathsf{D}K_N \big(\varphi'(x),\varphi(x)\big)=F_N(x)\big(\varphi(x)\big)^N-\mathsf{D}\tilde{K}_N \big(\varphi'(x),\varphi(x)\big)
\end{equation}
and  therefore, in order to recover~(i), it is is sufficient to choose~$J_N=K_N+\tilde{K}_N$. The function~$F_N(x)$
is independent on $\varphi$ and $\varphi'$ and is just given by the action of the linear operator $A_N$ on the 
coefficients of~$\mathsf{D}K_N$. 

A priori, we can start with any set of trial 
functions~$(a_n)_{n\in\{1,\dots,N\}}$ 
to build up our~$K_N$, then compute~$F_N$ by computing the~$N^\text{th}$ power of~$\hat{R}$.
Before we try to control the sign of~$F_N$ for even~$N$, 
the only restriction so far on the~$a$'s is to preserve~(ii): $a_n$ should
not increase faster than~$\varphi'^{N-n}\varphi^n$ at~$|x|\to\infty$. 
However we will now show that
our freedom is in fact restricted to the choice of one test function only.
 In other words, many different 
choices of $(a_n)_n$ will lead to the same~$F_N$ and therefore will not help to gain any piece of information
(in particular those leading to an identically vanishing~$F_N$). To put it very qualitatively, $\bar{\mathcal{Q}}_N$
is a very thin subspace in~$\mathcal{P}_N$ (of co-dimension~$N$ if seen as a vector space on smooth real functions)
and the kernel of~$D$ is too small (given by the solutions of a linear ordinary differential equation of order~$N$)
for~$D^{-1}\mathcal{Q}_N$ to decrease its codimension. 
To understand that, let us introduce 
the projector~$\hat{\Pi}_n$ on the~$n^\mathrm{th}$ component of~$a$. Our previous construction
of condition (i) can therefore be re-written
\begin{equation}\label{eq:ioperator}
  \hat{D}a=\hat{R}^N\hat{D}a\;.
\end{equation}
where $a$ is the element of~$\hat{\mathcal{P}}_N$ associated with the function~$J_N$. 
In section~\ref{subsec:closedalg}
we will show directly for small~$N$ that 
\begin{equation}\label{eq:RNDPi0}
  \hat{\mathsf{R}}^N\hat{D}(1-\hat{\Pi}_0)=0
\end{equation}

which has the following consequence: adding to $a$ any vector $b=(b_n)_n$ whose~$b_0=0$ will not affect
the left hand side of~\eqref{eq:ioperator} from which~$F_N$ is computed. Therefore~$F_N$ depends only
on one function, namely~$a_0$. All the others can be canceled without loss of generality.
Actually, if we start with~$a=(a_0,0,\dots,0)$, we have
$\tilde{a}\DEF\hat{D}a=(a'_0,Nva_0,0,\dots,0)=(\tilde{a}_0,\tilde{a}_1,0,\dots,0)$ and substituting
$\tilde{a}$ with~$a$ in \eqref{eq:AN} leads straightforwardly to~\eqref{eq:FN}.
If we start with~$a=(0,a_1,0,\dots,0)$, we have
$\tilde{a}\DEF\hat{D}a=(a_1,a'_1,(N-1)va_1,0,\dots,0)=(\tilde{a}_0,\tilde{a}_1,0,\dots,0)$ 
and~\eqref{eq:RNDPi0} can be (tediously) checked in the special cases~$N=2$, 3 and~4 : by substituting
$\tilde{a}$ with~$a$ in \eqref{eq:AN},  $\Lambda_N$ identically vanishes. The same remains true
for~$a=(0,0,a_2,0,\dots,0)$, ~$a=(0,0,0,a_3,0,\dots,0)$, etc. The next section
provides a systematic way of proving the last results and~\S~\ref{subsec:simplifiedstart} gives a more general argument.
 
\subsection{The closed algebra}\label{subsec:closedalg}

Let us prove~\eqref{eq:RNDPi0} by introducing an algebra of operators that may be represented by their action
on~$\hat{\mathcal{P}}_N$ or on the functions in~$\mathcal{P}_N$.
The operator~$S^{\downarrow}\DEF\varphi\partial_{\varphi'}$ when multiplied by~$v$
allow to formalise the substitution of~$\varphi''$ by~$v\varphi$. We also 
define~$S^{\uparrow}\DEF\varphi'\partial_{\varphi}$. The arrows recall that these operators 
raise and lower the component of~$a$:
\begin{equation}
  \hat{S}^{\downarrow}\begin{pmatrix} a_0\\a_1\\a_2\\
                                     \vdots\\a_{N-2}\\a_{N-1}\\a_N
                      \end{pmatrix}=
\begin{pmatrix} 0\\Na_0\\(N-1)a_1\\\vdots\\3a_{N-3}\\2a_{N-2}\\a_{N-1}
\end{pmatrix}\;;\qquad
  \hat{S}^{\uparrow}\begin{pmatrix} a_0\\a_1\\a_2\\
                                    \vdots\\a_{N-2}\\a_{N-1}\\a_N
                    \end{pmatrix}=
\begin{pmatrix} a_1\\2a_2\\3a_3\\ \vdots\\ (N-1)a_{N-1}\\Na_N\\0
\end{pmatrix}\;.
\end{equation}
If we introduce the diagonal matrices 
$\hat{\Lambda}_{\scriptscriptstyle\searrow}\DEF\diag(0,1,\dots,N)$,
$\hat{\Lambda}_{\scriptscriptstyle\nwarrow}\DEF\diag(N,\dots,1,0)$ and the raising and lowering operators $ \hat{S}^{\pm}$ defined by the following action on any $a$
\begin{equation}
  \hat{S}^{-}\begin{pmatrix} a_0\\a_1\\a_2\\
                                     \vdots\\a_{N-2}\\a_{N-1}\\a_N
                      \end{pmatrix}=
\begin{pmatrix} 0\\a_0\\a_1\\\vdots\\a_{N-3}\\a_{N-2}\\a_{N-1}
\end{pmatrix}\;;\qquad
  \hat{S}^{+}\begin{pmatrix} a_0\\a_1\\a_2\\
                                    \vdots\\a_{N-2}\\a_{N-1}\\a_N
                    \end{pmatrix}=
\begin{pmatrix} a_1\\a_2\\a_3\\ \vdots\\ a_{N-1}\\a_N\\0
\end{pmatrix}\;;
\end{equation}
We have $\hat{S}^{\downarrow}=\hat{S}^{-}\hat{\Lambda}_{\scriptscriptstyle\nwarrow}$,  $\hat{S}^{\uparrow}=\hat{S}^{+}\hat{\Lambda}_{\scriptscriptstyle\searrow}$.
Within this formalism, the operator~\eqref{eq:opD} is
\begin{equation}
  \hat{D}=\hat{S}^{\uparrow}+\hat{\partial}_x+v \hat{S}^{\downarrow}
         =\hat{S}^{+}\hat{\Lambda}_{\scriptscriptstyle\searrow}+\hat{\partial}_x+v\hat{S}^{-}\hat{\Lambda}_{\scriptscriptstyle\nwarrow}\;,
\end{equation}
with $\hat{\partial}_x\DEF\diag(\partial_x,\dots,\partial_x)$.
To implement the commutation rules between $\hat{S}^{\pm}$ and 
any diagonal matrix $\hat{\Lambda}=\diag(\lambda_0,\lambda_1,\dots,\lambda_N)$,
we define the diagonal matrices
\begin{equation}
  \hat{\Lambda}^{(1)}\DEF\diag(\lambda_1\lambda_2,\dots,\lambda_N,\lambda_0)\;;\quad
  \hat{\Lambda}^{(-1)}\DEF\diag(\lambda_N,\lambda_0\lambda_1,\dots,\lambda_{N-1})\;,
\end{equation}
and for any strictly positive integer $k$, we define
$\hat{\Lambda}^{(k)}\DEF(\cdots(\hat{\Lambda}^{(1)})^{(1)}\cdots)^{(1)}$,
$\hat{\Lambda}^{(-k)}\DEF(\cdots(\hat{\Lambda}^{(-1)})^{(-1)}\cdots)^{(-1)}$
where the superscript occurs $k$ times in the right hand side. We also
take $\hat{\Lambda}^{(0)}\DEF\hat{\Lambda}$. Therefore 
the $n^{\mathrm{th}}$ entry of
$\hat{\Lambda}^{(k)}$ for any integer $k$ is~$\big(\hat{\Lambda}^{(k)}\big)_{n,n}=\lambda_{n+k\ \mathrm{modulo}\  N+1}$.
Then we have
\begin{equation}\label{eq:SLambda_LambdaS}
   \hat{S}^{\pm}\hat{\Lambda}=\hat{\Lambda}^{(\pm1)}\hat{S}^{\pm}\;.
\end{equation}
In particular if we set~$\hat{\Pi}_{-1}=\hat{\Pi}_{N+1}=0$, for 
any~$n\in\{0,\dots,N\}$,
\begin{equation}\label{eq:SPi_PiS}
  \hat{S}^{\pm}\hat{\Pi}_n=\hat{\Pi}_{n\mp1}\hat{S}^{\pm}\;.
\end{equation}

We can also express the reduction operator~$\hat{R}$ in terms of the shift operators~$\hat{S}$. From~\eqref{eq:opR}, we have
\begin{equation}
  \hat{R}=\hat{\Pi}_N+\hat{A}+\hat{B}=\hat{\Pi}_N+\hat{C}\;.
\end{equation}
with
\begin{subequations}
\begin{eqnarray}
\hat{A}&\DEF&-\hat{S}^-\hat{\Lambda}_a\,\hat{\partial}_x
\;;\\
\hat{B}&\DEF&-(\hat{S}^-)^2\hat{\Lambda}_b\,v\;;
\end{eqnarray}
\end{subequations}
and 
\begin{subequations}
\begin{eqnarray}
\hat{\Lambda}_a&\DEF&\diag\left(1,\dots,\frac{1}{n+1},\dots,\frac{1}{N+1}\right)
\;;\\
\hat{\Lambda}_b&\DEF&\diag\left(N-1,\dots,\frac{N-n-1}{n+1},\dots,\frac{1}{N-1},0,\frac{-1}{N+1}\right)\;.
\end{eqnarray}
\end{subequations}
From \eqref{eq:SPi_PiS} we have~$\hat{S}^-\hat{\Pi}_N=0$ and therefore 
$\hat{C}\hat{\Pi}_N=0$. Besides, $\hat{C}^N$ has a unique non-zero element, 
namely~$(\hat{C}^N)_{N,0}$ and then $\hat{\Pi}_N\hat{C}^N=\hat{C}^N$. Therefore,  the
expansion
of $\hat{R}^N=(\hat{\Pi}_N+\hat{C})^N$ leads to
\begin{eqnarray}
  \hat{R}^N&=&\hat{\Pi}_N\left(1+C+\cdots+C^N\right)\;;\\
         &=&\hat{\Pi}_N\hspace{-2ex}\sum_{\substack{\mathrm{words}\;w=(l,m)\\0\leqslant |l|+|m|\leqslant N}}\hspace{-2ex}
        \hat{A}^{l_1}\hat{B}^{m_1}\hat{A}^{l_2}\hat{B}^{m_2}\cdots\;.
\end{eqnarray}
The last sum involves all the distinct 
words~$w$ that can be made with the two ``letters'' $\hat{A}$
and $\hat{B}$ whose length is $|l|+|m|$ between $0$ and $N$ (the zero-length
word is the identity). Each word~$w$ is uniquely defined by two multiple indices, i.e.
by two sequences $l=(l_1,l_2,\dots)$ and  $m=(m_1,m_2,\dots)$ of positive integers
 that all vanish above a finite rank. 
We define~$|l|\DEFt l_1+l_2+\cdots$  and~$|m|\DEFt m_1+m_2+\cdots$. Eventually,
the identity \eqref{eq:RNDPi0} holds if we can cancel, for each~$n\in\{1,\dots N\}$,
the product
\begin{equation}\label{eq:RNDPin}
  \hat{R}^N\hat{D}\hat{\Pi}_n=\hat{\Pi}_N\Big(\hspace{-2ex}\sum_{\substack{\mathrm{words}\;w=(l,m)\\0\leqslant |l|+|m|\leqslant N}}\hspace{-2ex}
        \hat{A}^{l_1}\hat{B}^{m_1}\hat{A}^{l_2}\hat{B}^{m_2}\cdots\Big)
         \big(\hat{\partial}_x+v\hat{S}^{-}\hat{\Lambda}_{\scriptscriptstyle\nwarrow}
         +\hat{S}^{+}\hat{\Lambda}_{\scriptscriptstyle\searrow}\big)\hat{\Pi}_n\;.
\end{equation} 
To compute this sum,  we start by rearranging each word in order to shift, with 
the help of
\eqref{eq:SLambda_LambdaS}, all
the~$\hat{S}^-$'s involved in~$\hat{A}$ and~$\hat{B}$ to the right:
\begin{equation}
  \hat{A}^{l_1}\hat{B}^{m_1}\hat{A}^{l_2}\hat{B}^{m_2}\cdots=\hat{\Lambda}_w\big(\hat{S}^{-}\big)^{|l|+2|m|}
\end{equation}
with the diagonal matrix
\begin{multline}
 \hat{\Lambda}_w= (-1)^{|l|+|m|}\underbrace{\hat{\Lambda}_a^{(-1)}
                      \cdots\hat{\Lambda}_a^{(-l_1)}}_{l_1\ \mathrm{factors}}\hat{\partial}^{l_1}_x
                     \underbrace{\hat{\Lambda}_b^{(-l_1-2)} \cdots\hat{\Lambda}_b^{(-l_1-2m_1)}}_{m_1\ \mathrm{factors}}
                     \,v^{m_1}\\
                     \underbrace{\hat{\Lambda}_a^{(-l_1-2m_1-1)}
                      \cdots\hat{\Lambda}_a^{(-l_1-2m_1-l_2)}}_{l_2\ \mathrm{factors}}\hat{\partial}^{l_2}_x
                     \underbrace{\hat{\Lambda}_b^{(-l_1-2m_1-l_2-2)} \cdots\hat{\Lambda}_b^{(-l_1-2m_1-l_2-2m_2)}}_{m_2\ \mathrm{factors}}
                     \,v^{m_2}
              \cdots 
\end{multline}
Now each word contributes to \eqref{eq:RNDPin} via three terms
\begin{subequations}\label{subeq:contribword}
\begin{eqnarray}
\hat{\Lambda}_w\hat{\Pi}_N\big(\hat{S}^{-}\big)^{|l|+2|m|-1}\hat{\Pi}_n
 \hat{S}^{-}\hat{S}^{+}\hat{\Lambda}_{\scriptscriptstyle\searrow}
&=&
\hat{\Lambda}_w\big(\hat{S}^{-}\big)^{|l|+2|m|-1}\hat{\Pi}_n\delta_{N-|l|-2|m|+1,n}
\hat{\Lambda}_{\scriptscriptstyle\searrow}\;;\label{eq:contribword_a}\\
\hat{\Lambda}_w\hat{\Pi}_N\big(\hat{S}^{-}\big)^{|l|+2|m|}\hat{\Pi}_n\hat{\partial}_x
&=&\hat{\Lambda}_w \big(\hat{S}^{-}\big)^{|l|+2|m|}\hat{\Pi}_n\delta_{N-|l|-2|m|,n}\hat{\partial}_x\;;\\
 \hat{\Lambda}_w\hat{\Pi}_N\big(\hat{S}^{-}\big)^{|l|+2|m|+1}\hat{\Pi}_n\,
 v\hat{\Lambda}_{\scriptscriptstyle\nwarrow}
&=&\hat{\Lambda}_w\big(\hat{S}^{-}\big)^{|l|+2|m|+1}\hat{\Pi}_n\delta_{N-|l|-2|m|-1,n}
\,v\hat{\Lambda}_{\scriptscriptstyle\nwarrow}\;.
\end{eqnarray}
\end{subequations}
The passage from the left to the right hand sides  has been obtained
by moving all the projectors to the right with~\eqref{eq:SPi_PiS}. The 
 Kronecker symbols come from~$\hat{\Pi}_r\hat{\Pi}_n=\delta_{r,n}\hat{\Pi}_n$.
 In \eqref{eq:contribword_a}, 
the diagonal matrix $\hat{S}^{-}\hat{S}^{+}=1-\hat{\Pi}_0$ can be  
replaced by~$1$ for~$n\neq0$. The left hand sides of~\eqref{subeq:contribword} contain
the same matrix $\big(\hat{S}^{-}\big)^{N-n}\hat{\Pi}_n$ whose all elements vanish
but one:~$\big(\big(\hat{S}^{-}\big)^{N-n}\hat{\Pi}_n\big)_{N,n}=1$. 
When acting on a 
vector~$a$, only $a_n$ gets involved together with the last
diagonal element  $\Lambda_{w,N}\DEFt\big(\hat{\Lambda}_w\big)_{N,N}$ of $\hat{\Lambda}_w$.  Using
 $\big(\hat{\Lambda}_{\scriptscriptstyle\nwarrow}\big)_{n,n}=N-n$ and
$\big(\hat{\Lambda}_{\scriptscriptstyle\searrow}\big)_{n,n}=n$, 
we eventually obtain a vector whose $\{0,1,\dots,N-1\}$ components vanish. The last one
being 
\begin{equation}\label{eq:RNDPina}
 \big(\hat{R}^N\hat{D}\hat{\Pi}_na\big)_N= \hspace{-3ex}\sum_{\substack{\mathrm{words}\;w=(l,m)\\0\leqslant |l|+|m|\leqslant N}}\hspace{-3ex}\Lambda_{w,N} \Big(
\delta_{N-|l|-2|m|+1,n}n\,a_n+
\delta_{N-|l|-2|m|,n}\partial_xa_n
+\delta_{N-|l|-2|m|-1,n}(N-n)\,v\,a_n
\Big).
\end{equation}
The presence of $\delta$'s   forces the relevant words we some on keep 
$|l|+2|m|\leqslant N-1$ since $n\geqslant 1$. Now we have 
\begin{equation}
  \Lambda_{w,N}=\alpha_{w,N}\, \partial_x^{l_1} v^{m_1}
\partial_x^{l_2} v^{m_2}\cdots
\end{equation}
 with the numerical coefficient
\begin{multline}
  \alpha_{w,N}=(-1)^{|l|+|m|}\frac{(N-l_1)!}{N!} 
                             \frac{(l_1+2m_1-1)!!}{(l_1-1)!!}\frac{(N-l_1-2m_1-1)!!}{(N-l_1-1)!!} \\
                             \frac{(N-l_1-2m_1-l_2)!}{N-l_1-2m_1!}
                             \frac{(l_1+2m_1+l_2+2m_2-1)!!}{(l_1+2m_1+l_2-1)!!}
                             \frac{(N-l_1-2m_1-l_2-2m_2-1)!!}{(N-l_1-2m_1-l_2-1)!!}\cdots
\end{multline}
is obtained because $(\hat{\Lambda}^{(-k)}_a)_{N,N}=1/(N-k+1)$ and  
$(\hat{\Lambda}^{(-k)}_b)_{N,N}=(k-1)/(N-k+1)$ for $0\geqslant k\geq N$.
We have also introduced the usual notation for any integer~$n$
\begin{equation}
  n!!\DEF\begin{cases} 2^p \, p! & \text{if\ } n=2p \text{ with $p$ integer;}\\[1em]
           \displaystyle \frac{(2p+1)!}{ 2^p \, p!} & \text{if\ } n=2p+1 \text{ with $p$ integer.} \end{cases}
\end{equation}
I have not been able to prove diectly that the right hand side of 
\eqref{eq:RNDPina} vanishes for an arbitrary $N$ and for any $n\in\{1,\dots,N\}$
but since we are now left with an explicit expression 
for a linear differential operator acting on~$a_n$, it can be checked, possibly
with the help of symbolic computations, for small specific~$N$. 
It is rather straightforward to prove
that~$\big(\hat{R}^N\hat{D}\hat{\Pi}_na\big)_N$ vanishes for
 $n=N$ and $n=N-1$  with arbitrary
$N$  and that it also vanishes for  $N\leqslant4$ and any~$n\in\{1,\dots,N\}$.

\subsection{Open questions for possible generalizations remaining with~$d=1$}

It would be interesting to see if the tools introduced above may be adapted 
for higher order linear differential equations. More challenging, and this will
constitute the object  future research, we may obtain some interesting information
on the solutions of some non-linear equations
like \eqref{eq:grosspitaevskii}. The price to pay is that the non-linear term 
prevents us to keep working in one~$\hat{\mathcal{P}}_N$ space only; then, we should
 work in the whole~$\bigoplus_N\hat{\mathcal{P}}_N$  and probably extend
condition (i) by factorising a strictly positive polynomial in $\varphi$ rather than
considering only~$\varphi^N$.

\section{Higher dimensional cases}\label{sec:anyd}

One natural generalization to dimension~$d$ is to consider the Schr\"odinger equation
\begin{equation}\label{eq:schrodingerd}
  \Delta_d\,\varphi=\sum_{\mu=1}^d\partial^2_\mu\varphi
=2\big(V-E\big)\varphi=v\varphi\
\end{equation}
where~$x=(x^\mu)_\mu\in\mathbf{R}^d$ (we keep definition \eqref{def:v}). In the following, Greek indices always
label the dimension and we will follow the usual convention of letting implicit the sum from $1$ to~$d$ over repeated
Greek indices unless the opposite is specified. 
 The partial derivative with respect to
the~$\mu^{\text{th}}$ coordinate is denoted by $\partial_\mu\DEF\partial/\partial {x^\mu}$.

We will work within the 
space $\hat{\mathcal{P}}_{N,d}\DEF\bigotimes_{N=1}^d\hat{\mathcal{P}}_N$ of 
smooth real functions~$(a_n)_n$ where~$n=(n_1,\dots,n_d)$ is a multi-index with~$n_\mu\in\{0,\dots,N\}$
 from which we can construct the set $\mathcal{P}_{N,d}$  of 
functions built from homogeneous polynomials, namely
having the form
\begin{equation}\label{eq:Padgeneral}
  P_a\big(\partial\varphi(x),\varphi(x),x\big)\DEF
\sum_n a_n(\partial\varphi)^{N-n}\varphi^{N(1-d)+|n|}{\ }_{\lvert_{x}}
\end{equation}
where~$|n|=n_1+\cdots+n_d$, $(\partial\varphi)^{N-n}$  stands  for
$(\partial_1\varphi)^{N-n_1}(\partial_2\varphi)^{N-n_2}\cdots
 (\partial_d\varphi)^{N-n_d}$.
To apply the same reasoning that led to gaps in the spectrum, condition~(i)
will be extended in~$d$ dimension by looking for a current~$J_N=(J_N^\mu)_\mu$ whose divergence
can be factorised by a $\varphi$-independent function times a positive function.
More precisely, each~$J_N^\mu$ is associated with an element of~$\hat{\mathcal{P}}_{N,d}$ and is constructed in order to
fulfill
\begin{equation}
  \mathrm{(i)}_d \qquad D_\mu\Big(J_N^\mu\big(\partial\varphi(x),\varphi(x),x,E\big)\Big)=\varphi^N F_N(x,E)\;.
\end{equation}
Then by integrating it on the whole~$\mathbf{R}^d$, provided that 
\begin{equation}
  \mathrm{(ii)}_d\qquad \int_{\mathcal{V}}D_\mu J_N^\mu\, \dmat^dx
= \int_{\partial\mathcal{V}} J_N^\mu \,\dmat^{d-1}\sigma_\mu\to0
\end{equation}
where $\mathcal{V}$ is a closed radius whose typical length~$R$ tend to infinity ($\dmat^{d-1} \sigma_\mu$ is the measure
on its boundary~$\partial\mathcal{V}$ whose surface growths algebraically with~$R$, therefore any exponential
decrease of $J_N$ will guarantee $\mathrm{(ii)}_d$), the condition that
$F_N(x,E)$ should change its sign for even~$N$ will hopefully lead to some constraints on~$E$.
The total derivative is defined as the linear operator in $\mathcal{P}_{N,d}$
\begin{equation}
   D_\mu\Big(P_a\big(\partial\varphi(x),\varphi(x),x\big)\Big)\DEF
   \frac{\partial P_a}{\partial(\partial_\nu\varphi)}\,\partial^2_{\nu,\mu}\varphi  
+\frac{\partial P_a}{\partial\varphi}\,\partial_\mu\varphi+
 \partial_\mu  P_a\;.
\end{equation}
When~$d>1$, we cannot get rid  of the second derivatives of~$\varphi$ as easily as for~$d=1$
because \eqref{eq:schrodingerd} provides us with only one 
substitution rule\footnote{Without further information on $V$. In the non-generic case of
a separable potential, there are in fact $d$ independent substitution rules. The substitution 
rules may also implement some symmetries if there are any, like in the case of integrable systems.}: it is only when 
grouped into a Laplacian, that the substitution $\sum_{\mu=1}^d\partial^2_\mu\varphi=v\varphi$
can be done. If we start looking for a~$J_N$ from a generic~$a$,
grouping the second derivative in~$D_\mu J_N^\mu$ into Laplacians will eventually impose some relations
on the~$(a_n)_n$. Following what we have explained in the previous section, we will however systematically
work up to a total derivative. We can also extend the factorisation in $\mathrm{(i)}_d$ to other functions
$\varphi$ and its derivative whose sign is fixed. Rather than~$\varphi^N$, we still can apply
the argument if we manage to obtain
\begin{equation}
  \mathrm{(i)'}_d \qquad D_\mu\Big(J_N^\mu\big(\partial\varphi(x),\varphi(x),x,E\big)\Big)
 =\Big(B_{N/2}\big(\partial\varphi(x),\varphi(x),x,E\big)\Big)^2 F_N(x,E)\;.
\end{equation}
for even~$N$ where~$B_{N/2}$ is an homogeneous polynomial in~$(\partial\varphi,\varphi)$ of 
degree~$N/2$.

\subsection{Attempt for $N=2$}\label{subsec:attemptN2}

Let us tentatively start with
\begin{equation}\label{eq:J2g}
  J_2^\mu=g^{\mu\nu\rho}_0\,\partial_\nu\,\varphi \partial_\rho\varphi+g^{\mu\nu}_1\,\partial_\nu\varphi+g_2^\mu
\end{equation}
where the~$g$'s are smooth functions of~$\varphi$, $x$ and $E$. We 
will start with
$d$ functions of type~$g_2$, $d^2$ functions of type~$g_1$ and $d^2(d+1)/2$
 functions of type~$g_0$ such that, without loss of generality,
\begin{equation}\label{eq:g0munurhosym}
  g^{\mu\nu\rho}_0=g^{\mu\rho\nu}_0\;.
\end{equation}
All the terms in $D_\mu J_2^\mu$ involving a second derivative in~$\varphi$ can be collected in
\begin{equation}
  (2g^{\mu\nu\rho}_0\,\partial_\rho\varphi+g^{\mu\nu}_1)\partial_{\mu\nu}^2\varphi\;.
\end{equation}
To construct a Laplacian, we must impose the parenthesis to be anti-symmetric
when~$\mu\neq\nu$ :
\begin{equation}\label{eq:g0munurhoantisym}
  g^{\mu\nu\rho}_0=-g^{\nu\mu\rho}_0\qquad(\mu\neq\nu)\;;
\end{equation}
\begin{equation}\label{eq:g1munuantisym}
  g^{\mu\nu}_1=-g^{\nu\mu}_1\qquad(\mu\neq\nu)\;,
\end{equation}
and independent of~$\mu$ when~$\nu=\mu$, 
 that is there are $d$ functions~$h^\rho_0$ and one function~$h_1$ such that
\begin{equation}
  g^{\mu\mu\rho}_0=h^\rho_0\qquad (\text{no summation on\ }\mu)
\end{equation}
and
\begin{equation}
  g^{\mu\mu}_1=h_1\qquad (\text{no summation on\ }\mu)\;.
\end{equation}
Combining~\eqref{eq:g0munurhosym} with~\eqref{eq:g0munurhoantisym} leads 
to~$g^{\mu\nu\rho}_0=g^{\mu\rho\nu}_0=-g^{\rho\mu\nu}_0=-g^{\rho\nu\mu}_0=g^{\nu\rho\mu}_0=g^{\nu\mu\rho}_0=-g^{\mu\nu\rho}_0$ and therefore~$g^{\mu\nu\rho}_0=0$ when~$(\mu,\nu,\rho)$ are pairwise distinct.
 Collecting in $D_\mu J_2^\mu$ the cubic terms in $\partial\varphi$ leads to
\begin{equation}
  (\partial_\varphi g^{\mu\nu\rho}_0)\,\partial_\mu\varphi\,\partial_\nu\,\varphi\, \partial_\rho\varphi=
  (\partial_\varphi h_0^\rho)\, \partial_\rho\varphi\sum_{\mu=1}^d(\partial_\mu\varphi)^2\;.
\end{equation}   
Therefore we will take
\begin{equation}\label{eq:dphih00}
  \partial_\varphi h_0^\rho=0\;.
\end{equation}
Quadratic terms in $\partial\varphi$ appearing in $D_\mu J_2^\mu$ are
\begin{equation}
  (\partial_\rho g^{\rho\mu\nu}_0+\partial_\varphi g^{\mu\nu}_1)\,\partial_\mu\varphi\,\partial_\nu\,\varphi
  =\sum_{\mu=1}^d\sum_{\nu=1}^d\partial_\mu h_0^\nu\,\partial_\mu\varphi\,\partial_\nu\,\varphi
   +\sum_{\mu=1}^d\sum_{\substack{\nu=1\\\nu\neq\mu}}^d \partial_\nu h_0^\mu\,\partial_\mu\varphi\,\partial_\nu\,\varphi
   +\sum_{\mu=1}^d \Big[(\partial_\mu\varphi)^2 (\partial_\varphi h_1-\sum_{\substack{\nu=1\\\nu\neq\mu}}^d\partial_\nu h_0^\nu)\Big]\;.
\end{equation}
Canceling each independent term requires
\begin{equation}
  \partial_\mu h_0^\nu+\partial_\nu h_0^\mu=0\qquad(\mu\neq\nu)
\end{equation}
and
\begin{equation}\label{eq:2moinsA}
  \partial_\mu h_0^\mu-\sum_{\substack{\nu=1\\\nu\neq\mu}}^d\partial_\nu h_0^\nu+\partial_\varphi h_1=0
\qquad (\text{no summation on\ }\mu)\;.
\end{equation}
The last equation appears as a  linear system of $d$ equations that can be rewritten with the help
of the $d\times d$ matrix $2-A_d$ where $(A_d)_{\mu,\nu}=1$. Then, 
$\det(2-A_d)=(-2)^{d-1}(d-2)$ and therefore
when~$d\neq2$,  it can be inverted and leads to
\begin{equation}\label{eq:dmuh0h1}
  \partial_\mu h_0^\mu=\frac{1}{d-2}\,\partial_\varphi h_1\qquad (\text{no summation on\ }\mu,\ d\neq2)\;.
\end{equation}
Now $h_0^\mu$ does not depend on~$\varphi$, by \eqref{eq:dphih00}, and then $\partial_\varphi h_1$ neither, hence
there exist two~$\varphi$ independent functions~$\tilde{h}_1$ and~$\tilde{\tilde{h}}_1$ such that
\begin{equation}\label{eq:h1tilde}
  h_1=\tilde{h}_1\varphi+\tilde{\tilde{h}}_1\;.
\end{equation} 
The relation \eqref{eq:dmuh0h1} implies (with now an implicit summation on $\mu$)
\begin{equation}\label{eq:h1tildeh0}
  \tilde{h}_1=\frac{d-2}{d}\partial_\mu h_0^\mu\;.
\end{equation}
In the special case~$d=2$, \eqref{eq:2moinsA} leads to
\begin{equation}\label{eq:cauchyd2}
  \partial_1 h_0^1=\partial_2 h_0^2
\end{equation}
and
\begin{equation}
\partial_\varphi h_1=0\qquad \;.
\end{equation}
Then we can keep \eqref{eq:h1tilde} together with \eqref{eq:h1tildeh0} even for~$d=2$.

Collecting all the previous relations we get
 \begin{equation}\label{eq:Dmuj2is}
   D_\mu J_2^\mu=\big(2h_0^\mu\Delta_d\,\varphi+
\partial_\nu\,g_1^{\nu\mu}+\partial_{\varphi}g_2^\mu\big)\partial_\mu\varphi+h_1\Delta_d\,\varphi+\partial_\mu g_2^\mu
\end{equation}
where of course, we can use the substitution~\eqref{eq:schrodingerd}.
Without loss of generality we can take~$g_1^{\nu\mu}=0$ for $\mu\neq\nu$ by possibly redefining
\begin{equation}
   g_2^\mu(\varphi,x,E)\mapsto g_2^\mu(\varphi,x,E)-\sum_{\substack{\nu=1\\\nu\neq\mu}}^d\int^\varphi_0 \partial_\nu\, g_1^{\nu\mu}(\varphi',x,E) \dmat\varphi'
\end{equation} 
since $\partial_\mu g_2^\mu\mapsto\partial_\mu g_2^\mu-\sum_{\substack{\nu=1\\\nu\neq\mu}}^d\int^\varphi_0 \partial_\mu\partial_\nu g_1^{\nu\mu}(\varphi',x,E) \dmat\varphi'=\partial_\mu g_2^\mu$ because the integrand cancels by~\eqref{eq:g1munuantisym}. To cancel
the parenthesis in~\eqref{eq:Dmuj2is} we must take
\begin{equation}
\partial_{\varphi}g_2^\mu=-\partial_\mu h_1-2h_0^\mu v\,\varphi
\end{equation}
The dependence in~$\varphi$ appears only through~\eqref{eq:h1tilde} and we can immediately integrate the last relation
\begin{equation}
  g_2^\mu=-\left(\frac{1}{2}\partial_\mu\,\tilde{h}_1+h_0^\mu v\right)\varphi^2
-(\partial_\mu\,\tilde{\tilde{h}}_1)\varphi+\tilde{g}_2^\mu
\end{equation}
where~$\tilde{g}_2$ is~$\varphi$-independent. With this expression, \eqref{eq:Dmuj2is} becomes
\begin{equation}
   D_\mu J_2^\mu=-\varphi^2\left[\frac{1}{2}\Delta_d\tilde{h}_1-v\tilde{h}_1+\partial_\mu(vh_0^\mu)\right]
   -\varphi[\Delta_d\tilde{\tilde{h}}_1-v\tilde{\tilde{h}}_1]+\partial_\mu\tilde{g}_2^\mu\;.
\end{equation}
The last term is irrelevant because it is a total divergence and can be reabsorbed in the definition of~$J_2^\mu$.
The second term does not contribute also since by integration by part it can be converted
to~ $\tilde{\tilde{h}}_1[\Delta_d-v]\varphi$ which vanishes.
with the use of~\eqref{eq:h1tildeh0}, the first term can be further 
simplified in order to keep~$h_0^\mu$ only. 

To sum up, condition~$\mathrm{(i)}_d$
can be obtained for~$N=2$ with 
\begin{equation}\label{eq:F2d}
  F_2=\frac{2-d}{2d}\,\Delta_d\partial_\mu h_0^\mu-\frac{2}{d}\,v\,\partial_\mu h_0^\mu
 -h_0^\mu\partial_\mu v
\end{equation}
with $h_0^\mu$ being any $d$ smooth functions such that
\begin{subequations}\label{eq:constraintsh0}
\begin{eqnarray}
  \partial_1 h_0^1&=&\partial_2 h_0^2=\cdots=\partial_d h_0^d\;;\\
  \partial_\mu h_0^\nu&=&-\partial_\nu h_0^\mu\qquad(\mu\neq\nu)\;.  
\end{eqnarray}
\end{subequations}
Our freedom of choosing~$J_2$ has therefore being reduced first because eliminating the second derivatives
of~$\varphi$ through its Laplacian impose severe constraints and second because, as in the~$d=1$ case, 
many different initial choices lead to the same~$F_2$; in others words the linear
application from~$a$ to~$F$ has a non-zero kernel.

For $d=1$, we have one test function~$h_0=g_0=a_0$ at our disposal but without any restriction on its derivative
and we can easily checked that, for~$d=1$, \eqref{eq:F2d}  is~\eqref{eq:F2}.

For $d=2$, we have~$F_2=-\partial_\mu(vh_0^\mu)$ and surprisingly \eqref{eq:constraintsh0} appears to be the 
Cauchy-Riemann conditions for~$h_0^1+\imat h_0^2$ to be analytic. Anyway, if we compute
\begin{equation}
  \int_\mathcal{S}F_2\,\dmat x\,\dmat y=\int_\mathcal{\partial S}v\big(\begin{smallmatrix}h_0^1\\h_0^2\end{smallmatrix}\big)\cdot \dmat\vec{\sigma}
\end{equation}
on a surface~$\mathcal{S}\in\mathbf{R}^2$ whose boundary~$\partial S$ coincide 
with the energy level~$V(x,y)=E$, that is~$v=0$,
the integrand of the right hand side vanishes ($\dmat\vec{\sigma}$ is an infinitesimal $2d$-vector normal 
to the curve~$\partial S$ pointing outwards say). Therefore whatever choice we make for~$h_0$,
for any energy~$E$ belonging to the image of~$V$ (where we know the spectrum lies), $F_2$ changes its sign and no information
can be obtained further.

For $d\geqslant3$, the constraints \eqref{eq:constraintsh0} are so strong that they limit
the choice of~$h_0$'s to polynomials in $x$ of degree at most~$3$. 
Indeed all the third derivatives of $h_0$ must cancel  (up to equation~\eqref{eq:partial3h}
included, any pair of
distinct Greek
letters denote any pair of distinct values of indices and no summation over repeated indices 
is left implicit\footnote{I am grateful to  Oleg Lisovyy\cite{Lisovyy11a} for providing the following
arguments that concisely and rigourously proved my first guess of \eqref{eq:h0quadratic}. }). First,
\begin{equation}
  \partial^2_{\mu\nu}h_0^\sigma=-\partial^2_{\sigma\nu}h_0^\mu
   =\partial^2_{\sigma\mu}h_0^\nu=-\partial^2_{\nu\mu}h_0^\sigma
\end{equation}
and therefore $\partial^2_{\mu\nu}h_0^\sigma=0$; then $\partial^3_{\mu\nu\rho}h_0^\sigma=0$
\begin{equation}
  \partial^3_{\mu\nu\rho}h_0^\sigma=0;\quad\partial^3_{\mu\nu\rho}h_0^\mu=0;\quad
\partial^3_{\mu\mu\nu}h_0^\rho=0.
\end{equation}
 Furthermore,
\begin{equation}
  \partial^3_{\mu\mu\mu}h_0^\nu=-\partial^3_{\mu\mu\nu}h_0^\mu
  =-\partial^3_{\mu\rho\nu}h_0^\rho=0
\end{equation} 
since for~$d\geq3$ we can always find an index~$\rho$ distinct from both $\mu$ and~$\nu$; 
eventually we have
\begin{equation}\label{eq:partial3h}
  \partial^3_{\mu\mu\mu}h_0^\mu=\partial^3_{\mu\mu\nu}h_0^\nu
      =-\partial^3_{\mu\nu\nu}h_0^\mu=-\partial^3_{\rho\nu\nu}h_0^\rho
      =\partial^3_{\rho\rho\nu}h_0^\nu=\partial^3_{\rho\rho\mu}h_0^\mu
      =-\partial^3_{\rho\mu\mu}h_0^\rho=-\partial^3_{\mu\mu\mu}h_0^\mu
\end{equation}
and all these third derivatives actually vanish as well. Now  $h_0(x)$ being a polynomial
of degree at most two in~$x$, the constraints~\eqref{eq:constraintsh0} on 
its coefficients leads to the general form
\begin{equation}\label{eq:h0quadratic}
  h_0^\mu(x)=h_0^\mu(0)+k x^\mu+{A^\mu}_\nu x^\nu-\frac{1}{2}\, l^\mu x^2+x^\mu\, l\cdot x
\end{equation}
where $A$ is a $d\times d$ constant antisymmetric real matrix, $k$
a real constant,
 $l$ a real
 constant  $d$-vector ; the Cartesian product~$l\cdot x =l^\mu x^\mu$ is used. 
Then, from~\eqref{eq:F2d} we get:
\begin{equation}\label{eq:F2dxElambda}
   F_2(x,E,\lambda)=4\big(E-V(x)\big)(k+l\cdot x)
  -\big(h_0^\mu(0)+k x^\mu+{A^\mu}_\nu x^\nu-\frac{1}{2}\, l^\mu x^2+x^\mu\, l\cdot x\big)\partial_\mu V(x)\;.
\end{equation}
Unlike what occurs for~$d=1$ where we are free to construct $F_2$
from a whole set of test functions~$x\mapsto a_0(x)$, for~$d\geq3$
 we are left with only 
$(d^2+3d+2)/2$  free $x$-independent parameters, 
namely~$\lambda=\big(h_0^\mu(0),k,{A^\mu}_\nu,l^\mu\big)$. 

Now the boundaries of the gaps must belong to the solutions of~\eqref{subeq:equationsforthegaps}.
The linearity of~$F_2$ in~$\lambda$ simplify considerably the computations. The 
conditions~\eqref{eq:Fdlambda} imply~\eqref{eq:Fdequals0} and are equivalent to
\begin{equation}
  E=V(x)\;;\quad \partial_x V=0\;.
\end{equation} 
Condition \eqref{eq:Fdx} is guaranteed if we choose for instance
\begin{equation}
  h_0^\mu(0)=-k x_c^\mu+{A^\mu}_\nu x^\nu+\frac{1}{2}\, l^\mu x_c^2+x_c^\mu\, l\cdot 
\end{equation}
and condition \eqref{eq:nondegenerate} is generically fulfilled.

Therefore, with our method,
 possible candidates for the gap boundaries are the critical points~$x_c$ of~$V$
which is not a surprise from a semi-classical point of view. With this method we cannot expect to find
more interesting and more relevant piece of information. Actually, some inequalities
concerning the global spectrum may be obtained if we are to maintain the sign 
of~\eqref{eq:F2dxElambda}, specially once a specific~$V$ is given; but our ambition was, as 
we have shown in \cite{Giacomini/Mouchet07a} for $d=1$, 
to obtain some local information in the very core of the spectrum.


\section{Ending remarks}
\subsection{Simplified starting point}\label{subsec:simplifiedstart}

For~$N=2$ and any~$d$ we have shown directly that without loss of generality we could have started with 
no term in~$\varphi$ in~$J_2^\mu$, that is with a current such that 
\begin{equation}
 \partial_\varphi J_2^\mu\big(\partial\varphi(x),\varphi(x),x,E\big)=0\;.
\end{equation}
Indeed, the~$h_0^\mu$ are~$\varphi$ independent, see \eqref{eq:dphih00}, 
and we could have taken~$\tilde{h}_1=0$ with no consequence on the result~\eqref{eq:F2d}. 
For $d=1$ and any~$N$, we have proven this result through \eqref{eq:RNDPi0}: we obtain all the possible~$F_N$'s
even if we restrict our self to~$a=(a_0,0,\cdots)$. This
result can be also obtained for any~$d$ and any $N$ in another way. First remark that if we start with
$J_N^\mu\big(\partial\varphi(x),\varphi(x),x,E\big)$
the second derivatives in 
\begin{equation}
  D_\mu J_N^\mu
 =\partial_\mu J_N^\mu+\partial_\varphi J_N^\mu\, \partial_\mu\varphi
 +\partial_{\partial_\nu \varphi} J_N^\mu\,\partial_{\mu\nu}\varphi
\end{equation}
can be eliminated with the help of~\eqref{eq:schrodingerd} if and only if
\begin{equation}\label{eq:dnuJmuL}
  \partial_{\partial_\nu \varphi} J_N^\mu=L\delta^\mu_\nu+{W^\mu}_\nu
\end{equation} 
 where $\delta$ is the Kronecker symbol, 
$W\big(\partial\varphi(x),\varphi(x),x,E\big)$ an antisymmetric~$d\times d$ matrix 
and~$L\big(\partial\varphi(x),\varphi(x),x,E\big)$ a function.
Then after the substitution of~\eqref{eq:schrodingerd}, an integration by part can be made
and we have
\begin{equation}
  D_\mu J_N^\mu
 =\partial_\mu J_N^\mu+Lv\varphi-D_\mu(\partial_\varphi J_N^\mu)\varphi+D_\mu(\partial_\varphi J_N^\mu)\;.
\end{equation}
By computing $D_\mu(\partial_\varphi J_N^\mu)$ in the same way, and iterating the process
up to infinity, we find that $D_\mu J_N^\mu$ can be written like
\begin{equation}
  D_\mu J_N^\mu=v\varphi\sum_{n=0}^\infty\frac{(-\varphi)^n}{n!}\partial_\varphi^n L + \sum_{n=0}^\infty\frac{(-\varphi)^n}{n!}\partial_\varphi^n \partial_\mu J_N^\mu+ D_\mu \left(J_N^\mu\varphi\sum_{n=0}^\infty\frac{(-\varphi)^n}{n!}\partial_\varphi^n J_N^\mu\right)\\;.
\end{equation} 
Up to a total derivative, starting from any~$J_N$, we therefore are always led to
\begin{equation}
   D_\mu \Big(\Pi_0 J_N^\mu\Big)=v\varphi \Pi_0L+\Pi_0\partial_\mu J_N^\mu
\end{equation}
 where~$\Pi_0$ in the evaluation at~$\varphi=0$ that can be expressed as
 \begin{equation}\label{eq:Pi0}
  \Pi_0=\sum_{n=0}^\infty\frac{(-\varphi)^n}{n!}\,\partial^n_\varphi
\end{equation}
for any analytic function of~$\varphi$. Therefore even if we start with a $J_N$ whose component takes the general form~\eqref{eq:Padgeneral}, working up to divergence terms, we will be led to the same identity
as if we had started with all the~$a_n$ such that~$N(1-d)+|n|>0$ being 
zero (this is consistent with~\eqref{eq:RNDPi0}).
In $N=2$, all the~$g$'s in~\eqref{eq:J2g} could have been taken independent of~$\varphi$ from the very beginning.

\subsection{Remark about condition $\mathrm{(i)}'_d$}

For $d\neq2$ and~$N=2$, condition $\mathrm{(i)}'_d$ with 
\begin{equation}
  B_1=b_1^\mu\partial_\mu\varphi+b_0\varphi
\end{equation}
where $\big(b_1=(b_1^\mu)_\mu, b_0\big)$ are $d+1$ smooth real functions of $x$,
does not  provide~$F_2$ explicitly but rather leads to a linear differential equation for it.
For instance, when $d=1$ we get
\begin{equation}\label{eq:F2d1iprimer}
  \frac{1}{2}\,b_1^2F_2'' +(2b^{}_1b_1'-b^{}_1b^{}_0)F_2'+\big[(b^{}_1b_1')'-vb_1^2-(b^{}_1b^{}_0)'+b_0^2\big]F_2
   =\frac{1}{2}a'''_0-2va_0'-v'a_0
\end{equation}
which reduces to \eqref{eq:F2} when we make the simplest choice $b_0=1$ and $b_1=0$.
For $d\geq1$, by reproducing the same line of reasoning as in \S~\ref{subsec:attemptN2},
we get a second order linear partial differential equation for~$F_2$
\begin{multline}\label{eq:F2d_iprime}
      \frac{1}{2}\,(b_1)^2\Delta_d F_2
      +\left(\frac{2}{d}\,b_1\cdot\partial_\mu b_1-b_1^\mu b_0\right)\partial_\mu F_2
      +\left[\frac{1}{2d}\Delta_d(b_1)^2-\frac{1}{d}\,v(b_1)^2-\partial_\mu(b_1^\mu b_0)+b_0^2
      \right]F_2\\
=\frac{2-d}{2d}\,\Delta_d\partial_\mu h_0^\mu-\frac{2}{d}\,v\,\partial_\mu h_0^\mu-h_0^\mu\partial_\mu v\;.
\end{multline}
instead of \eqref{eq:F2d} obtained for $b_0=1$ and $b_1=0$. 
The  constraints on $h_0^\mu$ now involve the $b$'s and are entangled with $F_2$:
\begin{subequations}
 \begin{eqnarray}
  \partial_1 h_0^1-\frac{1}{2}\,(b_1^1)^2F_2&
  =&\partial_2 h_0^2-\frac{1}{2}\,(b_1^2)^2F_2=\cdots=\partial_d h_0^d-\frac{1}{2}\,(b_1^d)^2F_2\;;
\label{eq:constraintsh0_iprime_i}\\
  \partial_\mu h_0^\nu+\partial_\nu h_0^\mu&=&b_1^\mu b_1^\nu F_2\qquad(\mu\neq\nu)\;.\label{eq:constraintsh0_iprime_ii}  
\end{eqnarray} 
\end{subequations}
For $d\geq1$, the only way to get rid of~$F_2$ from \eqref{eq:constraintsh0_iprime_i} is to 
take~$b_1^\mu$ independent of~$\mu$, and eventually~$b_1=0$ if we want
 \eqref{eq:constraintsh0_iprime_ii} not to involve~$F_2$ either. Then, since 
$F_2$ and $b_0^2F_2$ have the same sign, \eqref{eq:F2d_iprime} take us back to \eqref{eq:F2d}
that is to case~$\mathrm{(i)}_d$ where~$b_0=1$. 

Even for $d=1$ and a specific $v$, I did not exploit further these possibilities, but the choice
of $b$'s for which~\eqref{eq:F2d1iprimer} can be solved explicitly is rather limited (not to speak of
the control of the sign of its solutions). In any case of course,
we ought to work with simpler differential equations than the Schr\"odinger equation itself !

\subsection{Attempt for $d=2, $$N=4$}
The second remark concerns an attempt to obtain~$\mathrm{(i)}_d$ for~$d=2$ and~$N=4$. Using the same ideas as in the case
$N=2$ and with 
\begin{equation}
  J_4^\mu=\sum_{n=0}^4 g^\mu_{4-m,m}\,(\partial_1\varphi)^{4-n}\,(\partial_2\varphi)^n
\end{equation}
with, according to the last remark, taking~$g$ as~$\varphi$-independent. The constraints~\eqref{eq:dnuJmuL} imposed
by the elimination of the second derivatives of $\varphi$ in~$D_\mu J_4^\mu$ lead to two independent functions
instead of the ten~$g$'s. But if we go further to eliminate the cubic terms in~$\partial\varphi$, these functions must vanish 
identically and no non zero $g$ can be found this way. Extending~$\mathrm{(i)}_d$ to~$\mathrm{(i)}'_d$ with
\begin{equation}
  B_2=b_{20}\,(\partial_1\varphi)^2+b_{11}\,\partial_1\varphi\,\partial_2\,\varphi+b_{02}\,(\partial_2\varphi)^2+
    b_{10}\,\varphi\,\partial_1\varphi+b_{01}\,\varphi\,\partial_2\varphi+b_{00}\varphi^2
\end{equation}  
where the  $b$'s are smooth function of~$x$, leads to the same conclusion.

\section*{Acknowledgments} 
It is a pleasure to thank  Hector Giacomini and
Oleg Lisovyy for substantial contributions and stimulating discussions.

\end{document}